# An Accessible Method for Simulating Charged-Particle Optics, with Examples for Transmission Electron Microscopy

*Patrick McBean[1]\*, Zachary Milne[2], Arjun Kanthawar[2], Cameron O'Byrne[1], Khalid Hattar[2,3], Lewys Jones[1,4]*

[1]School of Physics, Trinity College Dublin, Dublin, Ireland

[2]Nanostructure Physics, Sandia National Laboratories, Albuquerque, NM 87123, United States

[3]Department of Nuclear Engineering, University of Tennessee, Knoxville, TN 37996, United States

[4]Advanced Microscopy Laboratory, Centre for Research on Adaptive Nanostructures and Nanodevices (CRANN), Dublin, Ireland

\* Corresponding author: mcbeanp@tcd.ie

## Abstract

The transmission electron microscope (TEM) has become an essential tool for innovation in nanoscience, material science, and biology. Despite these instruments being widely used across both industry and academia, academics may hesitate to propose substantial modifications to the optical setup due to the instrument's significant purchase price, fear of voiding the service contract, or downtime being unacceptable in shared user facilities. For instruments found in industry, similarly the risk-reward balance makes substantive modifications untenable. This limits the development of radically new optical geometries, and with the performance of the TEM largely being dictated by the specification of the objective lens pole-piece, exploring novel designs may be valuable.

Alternatively, potential lens designs can be analyzed rapidly and inexpensively using finite element analysis multiphysics simulation packages. Several are available, but here COMSOL Multiphysics was used, which is readily available in many universities. Changes to the geometry or materials of the lens can be investigated without any need to disassemble, reassemble, and realign the TEM column. Here we demonstrate an intuitive and accessible method to simulate charged particle optics using this 'digital twin' approach, with the hope that this encourages new creative and sustainable grassroots innovation in TEM lens design and microscope modification.

### Keywords

# Introduction

Transmission electron microscopes (TEMs) are used extensively in industry and research to explore the micron- to atomic-scale morphology, composition, and/or physiochemical landscape of materials (Lin et al., 2021). TEMs are the highest-resolution volumetric-sample microscopes because they probe sample volumes with high-energy (and thus smaller wavelength) electrons rather than lower-energy (larger wavelength) photons. The phrase "volumetric-sample" is used to demarcate from surface-probing techniques such as atomic force microscopy (AFM) and scanning tunneling microscopy (STM). In the TEM, a beam of electrons is formed by accelerating electrons through a high-voltage field, and then a series of lenses comprised of copper coils are used to focus the beam onto the sample. The basic TEM has grown into supporting revolutionary new imaging and experimental methods, such as aberration-corrected imaging (Hawkes, 2009; Rose, 2009), *in situ* TEM experimentation (Gai & Boyes, 2009; Zheng et al., 2015), and 4D Scanning Transmission Electron Microscopy (4D-STEM) (Zewail, 2014). This transforms the TEM from an imaging device into a full experimental platform. Although the TEM is essential for the progression of material science, nanoscience, and biology, the current commercial landscape of TEMs risks becoming an oligopoly dominated by a handful of large players, which limits the development of radically new optical geometries.

Since the final beam-shaping is performed in the objective lens, its construction dictates much of the performance specifications of the TEM, resulting in previous work discussing lens design (Award & Tsuno, 1999). While TEMs and other microscopes that probe samples with charged particles (charged-particle optics) share many characteristics with microscopes that probe samples with photons (optical microscopes, lasers, etc.), the former uses magnetic fields to modify the path of the particles, whereas the latter use glass lenses.

When a charged particle (such as an electron) passes through a magnetic field, it experiences a force $\boldsymbol{F}$ due to the Lorentz Force,

$$\boldsymbol{F} = q\boldsymbol{E} + q\boldsymbol{v} \times \boldsymbol{B} \tag{1}$$

where $\boldsymbol{v}$ is the velocity of the particle, and $\boldsymbol{E}$ and $\boldsymbol{B}$ are the electric and magnetic fields, respectively. This allows the particle beam to be manipulated using electromagnetic lenses, such as to focus the beam onto the sample. Due to the cross-product, there is a centripetal force experienced by the electron, so it moves on a spiraling, 3-dimensional path. For this reason, charged-particle optics devices are often more complex to model. At a practical engineering level, it is extremely difficult to integrate new lens designs into existing TEM platforms, in part because failure results in the loss of an expensive piece of equipment, but also due to structural and vacuum requirements, compounded by limited availability of space within the TEM column. The lens materials also tend to be expensive and unique to the companies who design them, which means development of new TEM electron optical designs can be very costly. Much of light microscopy on the other hand, uses standard dimensions and threads in mechanical components, with recent research exploring an open-hardware, modular design to suit a variety of imaging conditions (Courtney et al., 2020; Rosenegger et al., 2014).

Fortunately, modern computer programs have become sophisticated enough to model charged-particle optics, which allows researchers to model candidate designs before attempting to implement those changes in a real machine. At present, there are several commercial modelling programs – EOD (Zlámal & Lencová, 2011)**,** SIMION (Dahl et al., 1998), ANSYS (Gyimesi et al., 1999)**,** and Munro's (Munro et al., 2006) – however most of these programs are specialized and therefore are not normally used or taught in university engineering programs. COMSOL, the finite-element package which enables this methodology presented here, is ubiquitous in many colleges and universities. Therefore, most STEM graduates from the past 20+ years will likely have some familiarity with COMSOL and be able to quickly and easily pick up this methodology.

While this simulation methodology is simple enough to be accessible to students and non-specialist lens designers, it is powerful enough for serious designers to develop a 'digital twin' (Tao et al., 2019) of current electron optics for the development of new concepts. This could be from rapidly exploring small design modifications to whole new concepts of potential interest. We hope this encourages innovation outside of the main TEM manufacturers, leading to a more sustainable and growing field. To demonstrate its utility, we show how this methodology can model several of the cornerstone lenses used in modern TEMs and offer a detailed explanation of how the simulations are set up, executed, and analyzed.

**Theory**

Figure 1.A shows a cross-sectional schematic of a simplified TEM that embodies the key features of most TEMs, with the magnetic circuit shown in blue. Different sections of the TEM column

manipulate different aspects of the electron beam trajectory, including; the pre-sample or "probe-forming" sections (the filament, accelerating tube, and condenser lenses), the inter-sample or "beam-focusing" section (the pole-piece, sample exchange/positioning assembly, and objective lens), and the post-sample "projection and detection" sections (the projector lens and detector/camera). The innovations of aberration correctors mean that there are separate modules dedicated to aberration correction (Haider et al., 1995; Krivanek et al., 1997), but for brevity these are not considered here.

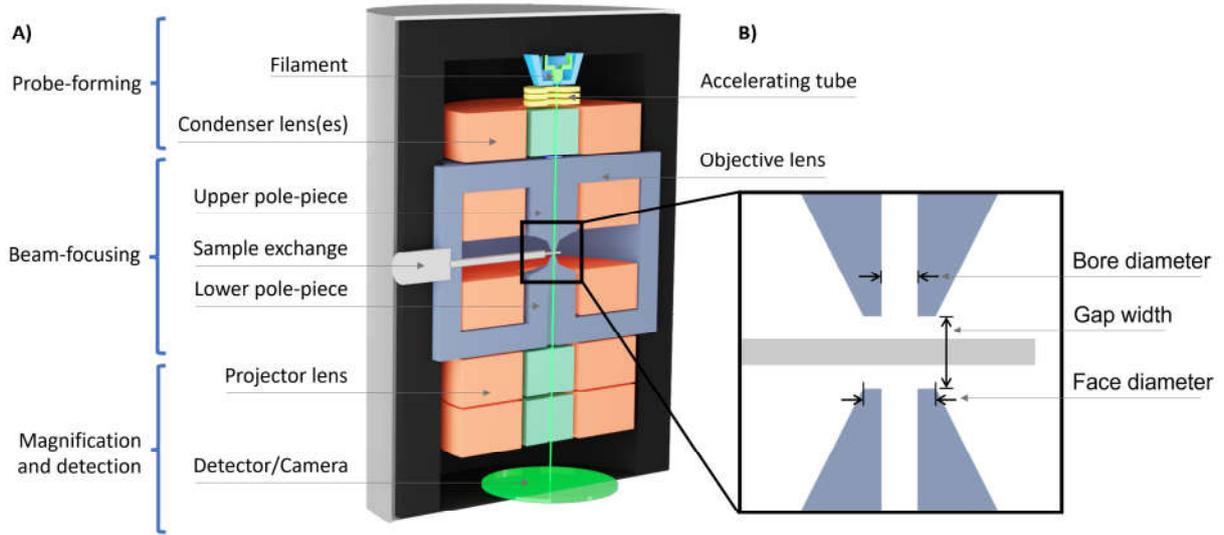

*Figure 1: A) Labelled cross-sectional view of a generic Transmission Electron Microscope (TEM). B) Close-up cross-sectional view of the pole-piece tips, with the important parameters of pole-piece design labelled.*

Magnetic lenses utilize the magnetic fields generated by passing a current through the lens coils in a TEM. The magnetization within the lens can be represented by two different equations, depending on whether the medium in question forms part of the magnetic circuit or not.

For the magnetic circuit, equation 2 shows the B-H curve relationship,

$$\boldsymbol{B} = f\big(||\boldsymbol{H}||\big)\frac{\boldsymbol{H}}{||\boldsymbol{H}||} \tag{2}$$

where $\boldsymbol{B}$ is the magnetic flux density, $\boldsymbol{H}$ is the magnetic field strength, and $f$ is the non-linear relationship between them, which is dependent on the material. For the remaining materials, the relative permeability given by the linear relationship

$$B = \mu_0 \mu_r H \qquad (3)$$

can be used, where $\mu_0$ it the magnetic permeability of free space, and $\mu_r$ is the relative permeability of the material.

Though the methodology discussed here can be applied to any magnetic lens used in a TEM, this article concentrates primarily on two different TEM lens systems that can be situated in the pole-piece section, which is shown in a closeup view in Figure 1.B. The specific lens being depicted here is an immersion lens (ImmL), the most common pole-piece design used in modern TEMs. Additionally, some research suggests that a snorkel lens (SnkL), in which the lower pole-piece is omitted, might provide the same resolution as an ImmL, but with an added benefit of more space for detectors and stimuli (Juma, Al-Nakeshli, et al., 1983; Juma, Khaliq, et al., 1983). To shed light on the potential of a SnkL, we apply our simulation methodology to a SnkL.

In an immersion style objective lens, the pre- and post-sample are both performance critical. The primary function of the pole-piece is to focus the electron probe on the sample. The important parameters of pole-piece designs that allow them to achieve tight and aberration-minimized focus are its bore diameter, gap width, and face diameter (Abbass & Nasser, 2012; Alamir, 2004; Ikuhara, 2002; Tsuno & Jefferson, 1998).

## Materials and Methods

In order to model the pole-piece, the 2D lens geometry needs to be designed and imported into COMSOL. We used SolidWorks for this, however, any software package capable of producing a DXF file could be used, or even COMSOL's inbuilt geometry editor. Due to the cylindrical symmetry present in the objective lens, this model can then be set up as a 2D axisymmetric model in COMSOL. The relevant material properties need to be assigned to each domain of the geometry for COMSOL to correctly calculate the propagation of the magnetic fields within the lens, and these can be selected either from COMSOL's existing library, or manually entered. If designing new parts, specifications directly from the suppliers can be used to ensure the simulation is accurate to the exact alloys being used. Additionally, the correct constitutive relationship needs to be specified for each domain, from equations 2 and 3. Components of the magnetic circuit are assigned the former, using the B-H curve, and the non-magnetic elements use the latter, the relative magnetic permeability. Next, the current is applied to the excitation coils in order to generate the magnetic field. This defines the excitation of the lens coils in ampere-turns (At). Finally, the model needs to be meshed to a suitable fineness depending on the solution accuracy required. Once ran, the 2D solution to the magnetic fields can be wrapped to produce a 3D result.

The use of a cylindrically symmetrical system greatly reduces the computational requirements of the simulation and simplifies the process of creating the geometry. However, geometries which are not azimuthally symmetric can be investigated also, by creating a 3D geometry and running a full 3D simulation in much the same way as for a 2D axisymmetric system.

A more detailed methodology is available in the supplemental information.

## Results & Discussion

To demonstrate its utility, we chose two distinct and relevant magnetic lens designs to use with the simulation methodology. These designs are the immersion lens (ImmL), and the snorkel lens (SnkL). Each of these is already used in either scanning electron microscopes or TEMs, so the results will be relevant to all forms of TEM. Despite the relatively simple examples shown here, there is room for enhanced and/or specialized capabilities through more complex designs.

### Immersion Lens (ImmL)

The immersion lens (ImmL) is the lens design used in the vast majority of TEMs today. Figure 2.A shows the ImmL assembly. The ImmL design has a rich history, starting with Ernst Ruska's original conception of a double pole-piece (Ruska, 1934), to the first of what could be considered the "modern" design shrouded by a magnetically soft iron yoke pioneered by von Ardenne (von Ardenne, 1944). The soft iron yoke was conceived as a means of controlling stray fields, while the double pole-piece was designed to be symmetric to best direct the magnetic flux towards the sample while maintaining parallel flux lines from face to face. This provides optimal focusing for the electrons. Another benefit of this design is that it allows the user to load a sample into the gap with relative ease, since the path of travel from the exterior to the interior rest position (in between the pole-piece faces) is unobstructed.

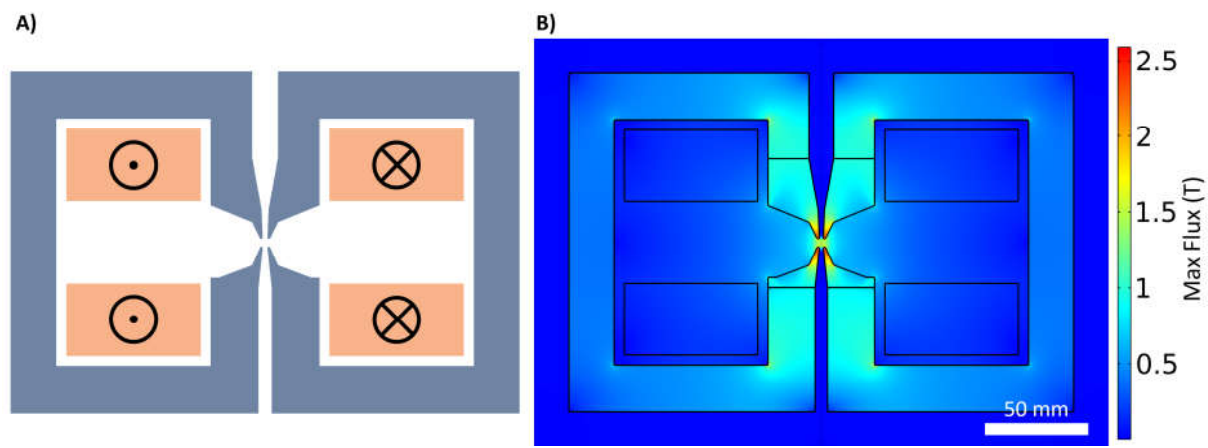

*Figure 2:A) The immersion lens in a Transmission Electron Microscope. The X indicates into-plane coil windings, whereas the • indicates out-of-plane coil windings. B) Magnetic field heat map, showing the flux concentrating in the pole-piece tips.*

The magnetic field for the ImmL is generated by annular solenoids known as coil windings, highlighted in orange in Figure 2.A. In real designs, the location and shape of the coils may be more complex than the design illustrated here, but their function is the same. For some manufacturers, there may be one coil above the pole gap and one below, or both may be on one side. Often water cooling is integrated next to or between the coils to keep the excitation current stable. The relevant parameters of the coils for objective lens design are: 1) the number of turns, N; 2) their length, L; and 3) the DC current which is applied to them, A. The field produced by the solenoids is focused into the gap where the sample would reside by the pole-pieces.

**Snorkel Lens (SnkL)**

The snorkel lens (SnkL) is sometimes referred to as a "pancake" lens (Williams & Carter, 2009). Figure 3.A shows a design of a SnkL and 3.B shows the distribution of magnetic flux density in this lens when active.

A potential benefit of removing the lower pole-piece of the ImmL, transforming it into a SnkL, is that it would make space for more sensors or stimuli, which would greatly benefit the field of in situ TEM characterization since 1) more sensors provide more information about the structure and composition of the sample and its interactions with stimuli, and 2) more stimuli greatly expand the possible physical phenomena that can be explored. However, the performance of the SnkL will be affected by the removal of the lower pole-piece due to reduced focusing power and increased aberrations. A SnkL less closely resembles the 'thin lens' approximation (Fultz & Howe, 2008) than an ImmL, so it is unclear whether a SnkL alone can provide enough focusing power to compete with the resolution attained by the ImmL.

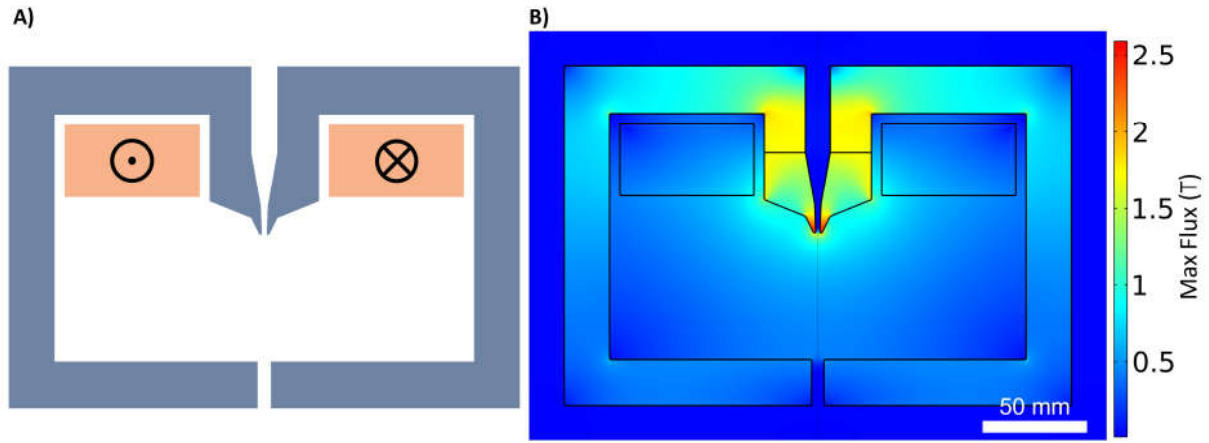

*Figure 3: A) An example of a snorkel lens. The X indicates into-plane coil windings, whereas the • indicates out-of-plane coil windings. B) Magnetic field heat map. Due to the higher coil excitation required to reach saturation, higher levels of flux are present in the upper region of the pole-piece.*

**Magnetic Flux Curves**

Through plotting the magnetic flux intensity along the optical axis, as shown in Figure 4.A, we can examine the profile of the magnetic field along the beamline for each lens type. While the immersion lens is close to symmetrical, and both lenses perform similarly in the region of the top pole, the asymmetry of the snorkel lens causes a tail where the magnetic field expands into free space without the lower pole to confine it. Despite the snorkel lens design requiring much higher excitation to achieve saturation, it still achieves a lower peak field intensity. However, there is more experimental space available in the SnkL, allowing for more exotic stages or in situ work.

In the case of an ideal lens, the flux curve would be a delta function, with no width. As the lens deviates from the thin lens approximation, the full width at half maximum (FWHM) increases due to aberrations. These flux curves can be used to analytically determine these aberrations, although this often requires closed-source proprietary software.

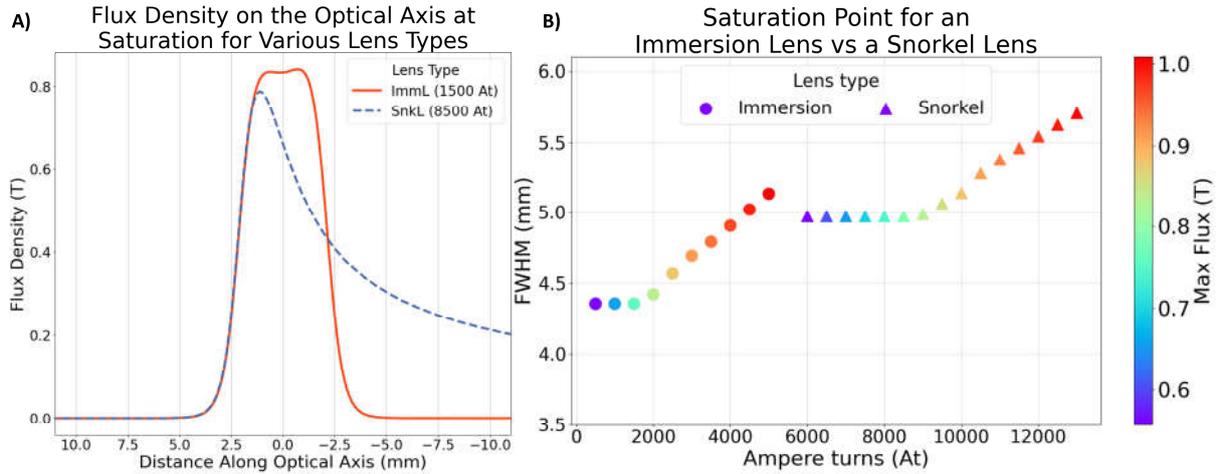

*Figure 4: Analysis of the magnetic flux curves along the optical axis. A) The magnetic flux density along the optical axis for an immersion lens (red, solid) compared to that of a snorkel lens (blue, dashed). Each are at their saturation point, determined by B. B) The saturation point of each lens, where the FWHM starts to increase once a threshold excitation is exceeded.*

For both types of lenses, the FWHM of the flux curves was calculated for a range of different lens excitations (see Figure 4.B). This relationship is characteristically flat for lower excitations, where increasing the excitation increases the magnetic flux without affecting the FWHM, and linear once the saturation point is reached. Once the saturation point of the alloy used in the pole-piece is reached, no more flux can pass through it, resulting in the remaining flux "spilling out" of the poles, widening the FWHM. The saturation point is therefore this "knee" point, just before the FWHM begins to increase linearly. Therefore, the optimal excitation to use is the maximum excitation that doesn't increase the FWHM.

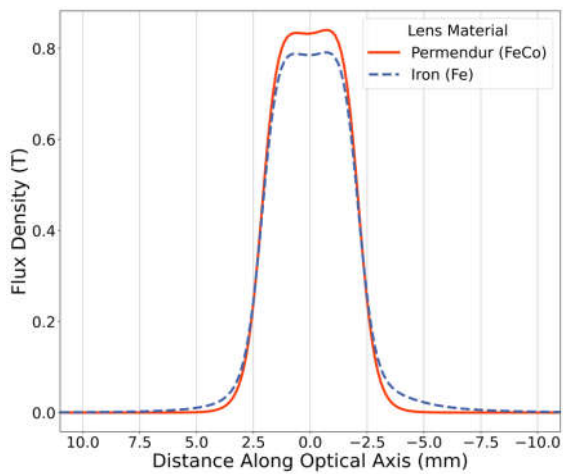

*Figure 5: Comparison of permendur (red, solid) and iron (blue, dashed) as the material for the pole-piece. Permendur's higher maximum saturation point results in a higher peak flux intensity at the same excitation (1500 At).*

An iron-cobalt alloy named permendur is usually chosen for pole-piece manufacturing due to its high magnetic saturation value (Tsuno & Jefferson, 1998). Pure iron is cheaper, but Figure 5 demonstrates the reduced performance of the lens, due to achieving a lower maximum flux density at the same lens excitation. Raising the excitation to compensate for this will result in widening of the peak, resulting in increased aberrations.

## Conclusions

The methodology presented here creates a platform for an intuitive and accessible way to simulate charged particle optics utilizing a 'digital twin' concept. Using the methodology presented here, each element can be refined one at a time, building up a more complex model of an instrument. Further work can extend the modelling to multi-lens systems, or fully leverage it to create digital twins of extant microscopes. Indeed, a complete digital twin would allow researchers the freedom to test design ideas, which would otherwise be costly and time consuming. Through constructing simulations for two lens systems; a double pole-piece and a single pole-piece, the saturation points of the double vs single pole-piece systems could be determined and compared, along with producing magnetic flux curves along the beamline which can be used to analytically determine lens aberration values. This methodology will be useful for novices to gain a richer understanding of SEM and TEM imaging conditions and for experts looking to push the limits in novel electron optic concepts. We hope that making these tools more open reduces the barrier to entry and encourages more sustainable innovation.

## Acknowledgements

LJ would like to acknowledge SFI/Royal Society & PMB/COB would like to acknowledge SoP/AMBER. All authors would like to acknowledge technical insights provided by Dr. Doug Medlin, Victor Chavez, and E. Ted Winrow at Sandia National Laboratories. They would also like to acknowledge Dr Katherine Jungjohann for helpful discussions. This work was performed, in part, at the Center for Integrated Nanotechnologies, an Office of Science User Facility operated for the U.S. Department of Energy (DOE) Office of Science. Sandia National Laboratories is a multimission laboratory managed and operated by National Technology & Engineering Solutions of Sandia, LLC, a wholly owned subsidiary of Honeywell International, Inc., for the U.S. DOE's National Nuclear Security Administration under contract DE-NA-0003525. The views expressed in the article do not necessarily represent the views of the U.S. DOE or the United States Government.

## Competing Interests.

Competing interests: The authors declare none.

# Supplementary Info - Simulation Methodology

This simulation methodology employs the COMSOL Multiphysics finite-element modelling software developed in Stockholm, Sweden. We have included an example simulation [1], which may be used for learning or to modify a particular application.

- Creating and exporting the CAD geometry

As the microscope column can be approximated as being cylindrically symmetrical, significant performance gains can be made by performing a 2D simulation of the magnetic fields, and then wrapping the solution to form a 3D solution which can then be used for further simulations (such as charged particle tracing). The geometry used in the COMSOL simulations was created in SolidWorks (3DS, Vélizy-Villacoublay, France, E.U.) as five distinct CAD parts: the top and bottom pole-pieces, the upper and lower solenoids, and the remainder of the magnetic loop. A 2D drawing of the assembly is shown in Figure 1 and was saved as a .DXF file. While this geometry is a simple one, one can put as much detail into each lens and lens assembly as desired being aware of the trade-off in computational resources needed.

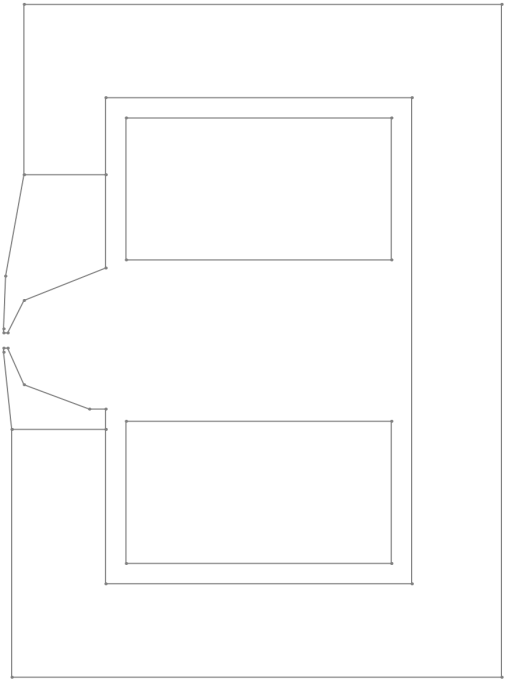

*Figure 1: The CAD geometry drawn in SolidWorks and exported as a .DXF file.*

- Importing the CAD geometry to COMSOL

First, a 2D axisymmetric component was added to a blank COMSOL model. Next, the geometry was imported using the *import* feature in the *geometry* node. Sharp corners can be problematic for the simulation, so a small fillet (0.2mm) was added to each corner. Finally, an environment part was added by specifying a *rectangle* encompassing the existing geometry (see Figure 2). This environment part can be created in COMSOL using COMSOL's simple-geometry tools (i.e. as a rectangle) instead of saving and importing it with the lens assembly because COMSOL can have difficulty identifying individual parts that have been imported when they intersect another part (e.g. the environment part with the lens part). *Build All Objects* will then finalise the geometry, and raise alerts if there are any errors or warnings. Note that, depending on the CAD design, it may be necessary to save and import the five parts individually, rather than as a single sketch, since COMSOL may incorrectly identify individual parts when the parts are imported as a single sketch.

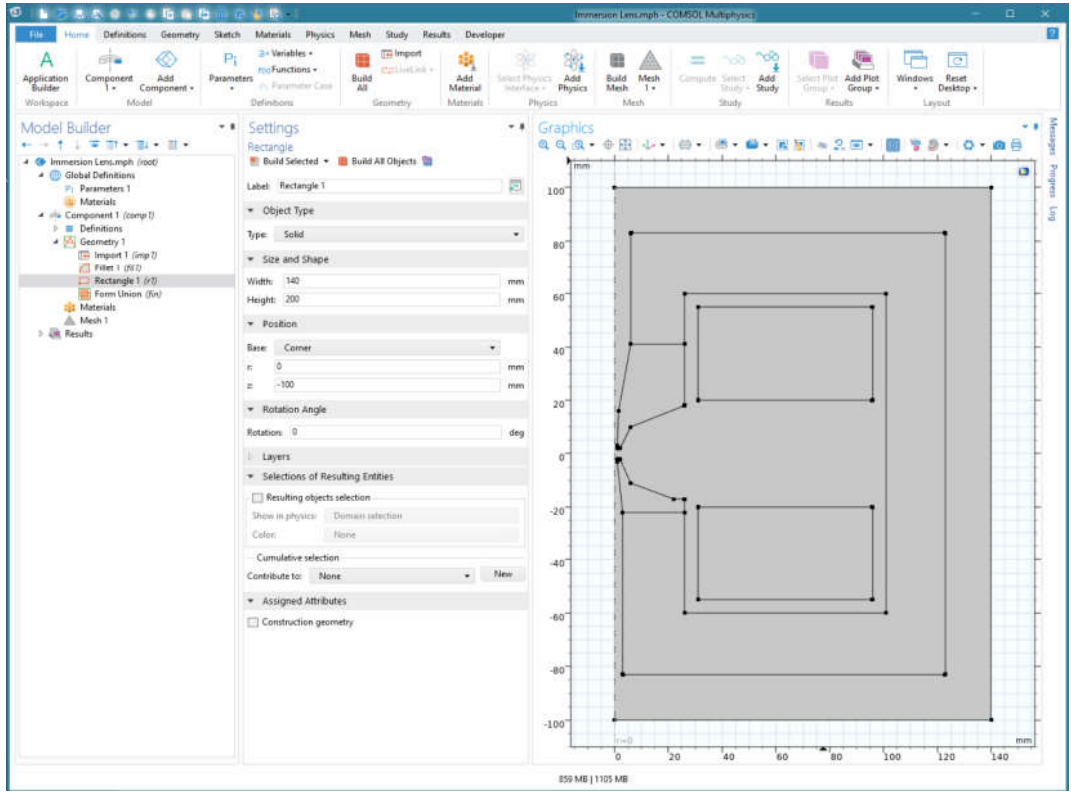

*Figure 2: Importing the geometry into the COMSOL simulation.*

- Defining magnetic and non-magnetic materials

Next, the materials were defined. These were added through the *add a new material* in the *materials* node. Recent versions of COMSOL group materials by application, but they can also be found using the search function. Using this, *air*, *soft iron*, *supermendur*, and *copper* were added.

To assign a *material*, click it in the node menu and then select the *domain* to which the material applies. *Supermendur* was assigned to the pole-piece assembly, *soft iron* to the remainder of the magnetic circuit, *copper* to the coils, and *air* to the environment. Under the *soft iron* or *supermendur* material nodes, it shows the definition of the B-H curve, which comes from a table of B (magnetic flux density magnitude) and H (magnetic field magnitude) values. This table can be modified to create any B-H curve desired, as long as the curve has an inverse function (i.e. is strictly monotonic). For example, values from a metal supplier can be substituted into the simulation to obtain more accurate results. Specifications for the copper can also be updated as desired (see Figure 3).

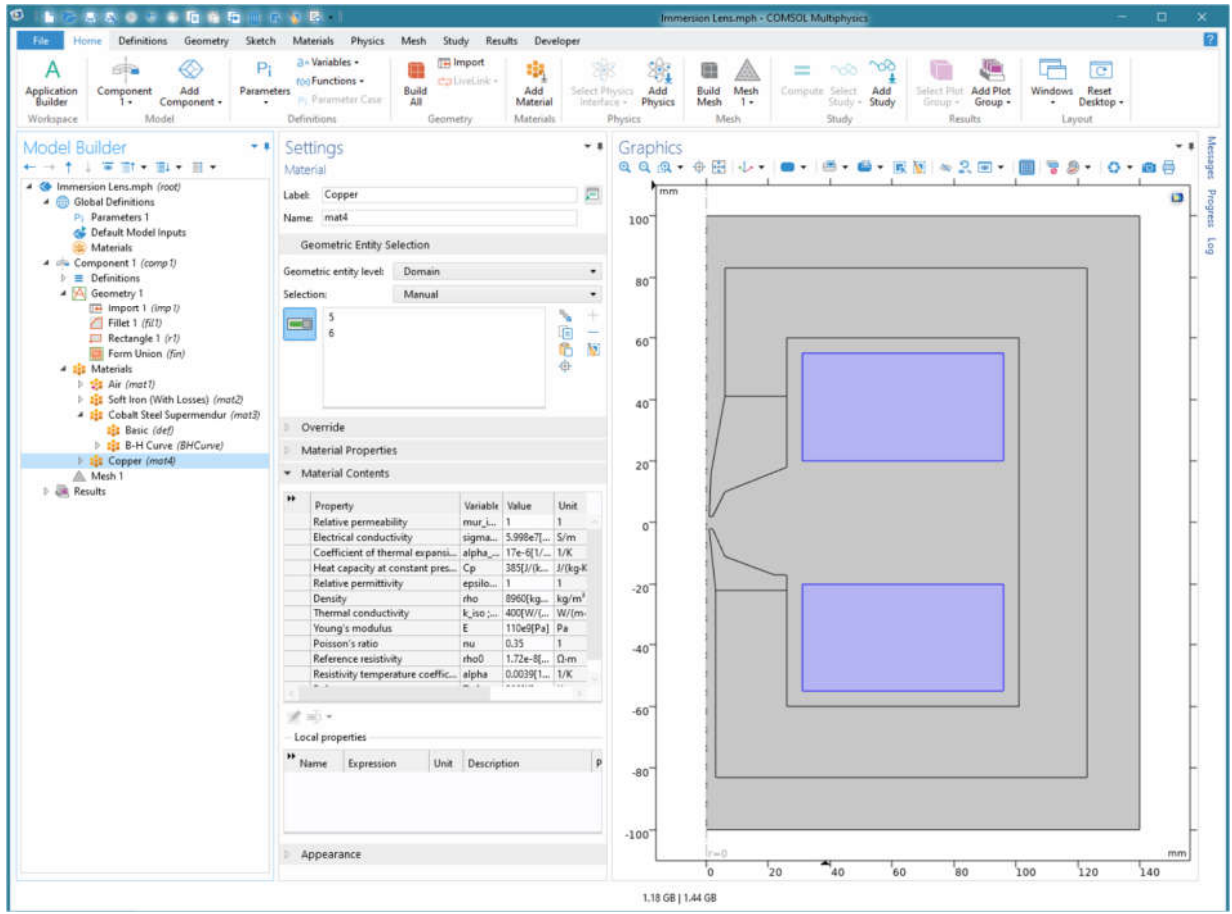

*Figure 3: Assigning materials to each domain of the geometry.*

- Adding the physics interface

The *magnetic fields (mf)* physics interface was then added to the simulation. While this adds some default nodes, several extras need to be added. A second *Ampère's Law* node was added, under which the *constitutive relation B-H* was changed to *B-H curve*. The relevant domains (the poles and the magnetic loop) were selected for this law, so that in these domains, their B-H curves are used due to their nonlinear behavior. The remaining domains will automatically use the default *Ampère's Law* node, which uses the relative permeability of the material for the constitutive relation.

There are two ways to add the excitation to the coils, either by directly adding a *coil* node to the *magnetic fields* interface, or by adding an *external current density* instead. In the former case, the *conductor model* must be changed to *homogenized multiturn*. The *current* and *number of turns* can then be specified. These can also be added as variables at the start of the model under *Parameters* for ease of use. Finally, the two coil domains are selected for this node.

- Creating the mesh

The *mesh* settings in COMSOL are both user friendly (for occasions where only a simple mesh is required) and flexible where more complicated meshes are necessary. The appropriate meshing fidelity should always be determined by running the simulation at decreasing mesh sizes until the results no longer differ to the desired degree of fidelity, but for the sake of simplicity, in this example the finest default mesh *(extremely fine)* was used.

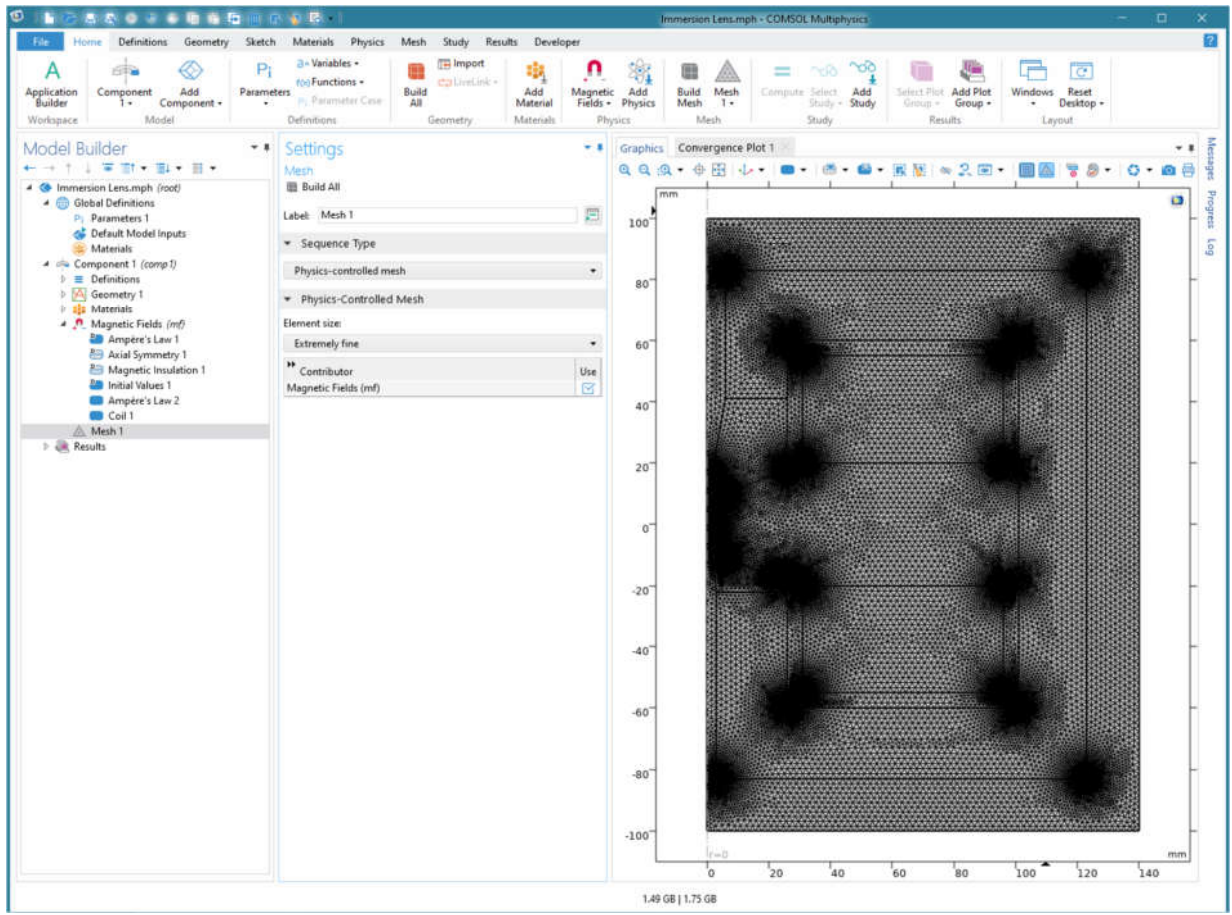

- Running the study

Finally, the study type must be specified. Here, a *stationary* (time-independent) study was added, and subsequently ran. Even with the fine meshing, this only took several seconds on a midrange CPU.

- Data analysis

COMSOL produces a set of default plots which can be modified to the needs of the user. To plot the flux along the optical axis, a new *dataset* is created, of the type *cut line 2D*. The start and end points of the line are specified (in this example, r=0, and z=(-10 →10)). A *1D plot group* is then added, to which a *line graph* is added, and the previously created *cut line 2D* dataset is specified for it (see Figure 5). The rest of the values are automatically populated, but can be tweaked with if required. Alternatively, the data can be exported for post-processing using other software.

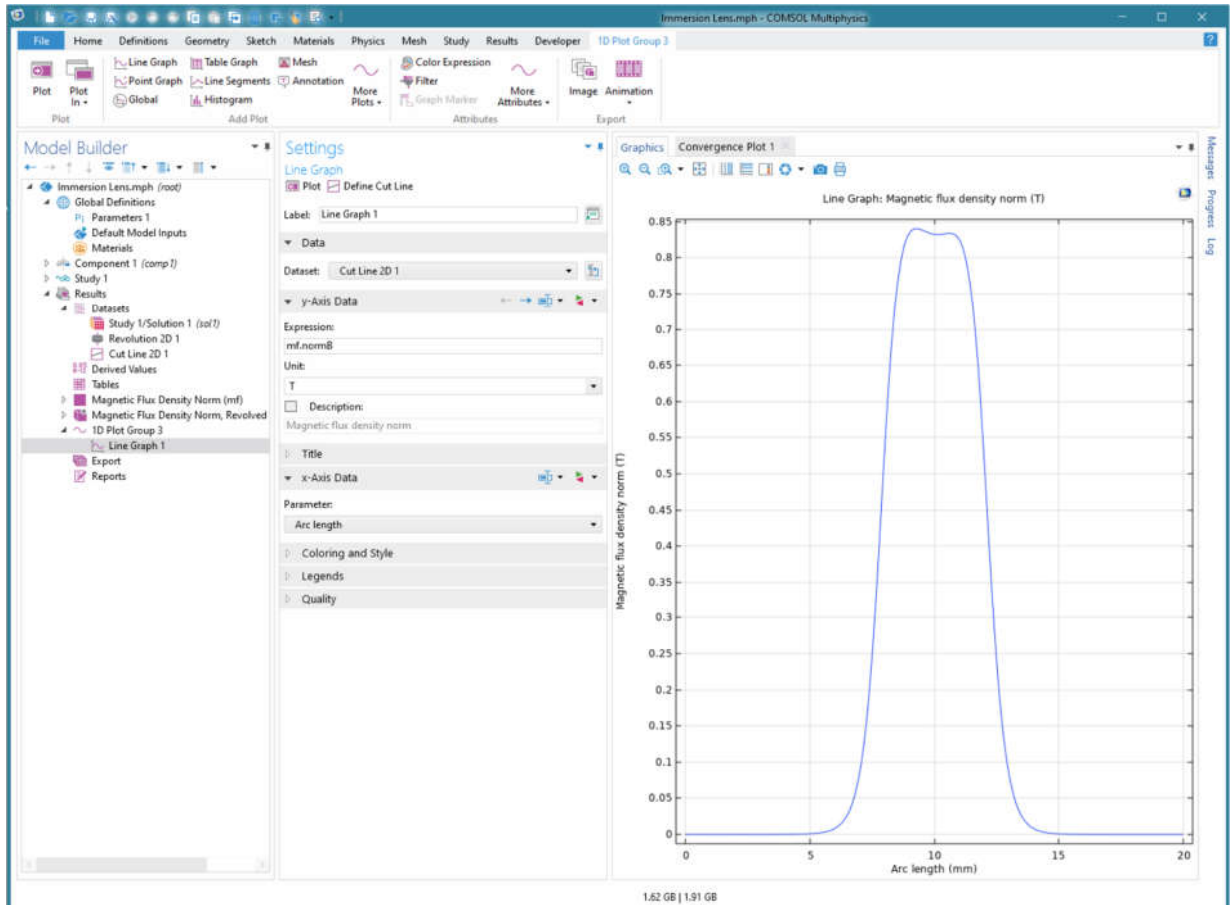

*Figure 5: Plotting the magnetic flux density along the optical axis.*

- 3D modelling

For models which are not azimuthally symmetrical, a full 3D simulation can be performed in much the same way, by adding a 3D component at the start instead of a 2D axisymmetric one. The imported geometry will require also being 3D (such as .STL files), or alternatively COMSOL's inbuilt geometry editor can be used. Performing a 3D simulation requires significantly more computational power, but careful refinements to the meshing level can reduce this.